\documentclass[final,5p,times,twocolumn]{elsarticle} 

\usepackage{graphicx}

\usepackage{amsmath}   
\usepackage{xcolor}
\usepackage{caption}
\usepackage{subcaption}
\usepackage{lineno}

\begin{document}
\begin{frontmatter}

\title{Assessment of the potential of SiPM-based systems for bioluminescence detection}
\author[add1]{S.Lomazzi\corref{cor}}
\ead{s.lomazzi@uninsubria.it}
\author[add1,add2]{M.Caccia}
\author[add3]{C.Distasi}
\author[add3]{M.Dionisi}
\author[add3]{D. Lim}
\author[add1,add4]{A. Martemiyanov}
\author[add1]{L.Nardo}
\author[add3]{F.A. Ruffinatti}
\author[add1,add2]{R.Santoro}
\cortext[cor]{Corresponding author}
\address[add1]{Universit\`a degli Studi dell'Insubria, Como, Italy}
\address[add2]{INFN, Milan, Italy}
\address[add3]{Universit\`a degli Studi del Piemonte Orientale, Novara, Italy}
\address[add4]{ITEP, Moscow, Russia}
\begin{abstract}
Bioluminescence detection requires single-photon sensitivity, extremely low detection limits and wide dynamic range. Such performances were traditionally assured by photomultiplier-tubes based systems. However, development of novel applications and industrialisation call for the introduction of more robust, compact and scalable devices. Silicon photomultipliers were recently put forward as the alternative to phototubes for a new generation of flexible and user friendly instruments. In this article, the figures of merit of a silicon-photomultiplier based system relying on a compact, low cost system are investigated. Possible implementations are proposed and a proof-of-principle bioluminescence measurement is performed.
\end{abstract}

\end{frontmatter}
\section{Introduction}
Bioluminescence techniques have known a widespread diffusion in the last decades, and are nowadays routinely applied in several different fields of biophysics, biochemistry and biomedicine \cite{biolumin} \cite{chemiolumin}. The availability of on-purpose bioluminescence kits for the quantification of a wealth of metabolites has shifted the range of application from academy to industry, increasing the panel of potential users and calling for the development of simpler, stealthier and more flexible detection systems. On the other hand, bioluminescence detection remains challenging inasmuch as the combination of bioluminescent emission properties and the concentration range of metabolites to be quantified sets stringent requirements on the detection system. Namely, single-photon sensitivity, the ability of discriminating from noise very dim light signals (low detection limit) and wide dynamic range are critical for any luminometry device. Until very recently such performances could only be assured by apparatuses relying on photomultiplier tubes (PMTs) as the detectors. However, this class of sensors is cumbersome, bulky and fragile. High voltage operation is required and the measurements are hindered by the presence of magnetic fields. These features severely limit a full exploitation of PMT-based luminometers in industrial environments and diagnostics laboratories, as well as the further development of in loco assays, e.g. for the assessment of water quality or applications in veterinary.

Silicon photomultipliers (SiPMs) \cite{overview} are multi-pixel solid-state detectors. Basically, they consist in arrays of single-photon avalanche diodes (SPADs) biased above the breakdown voltage and operated in the Geiger-Muller regime. The independently operating cells are connected to the same readout line and the combined output signal corresponds to the sum of all fired cell signals. Thus, SiPMs combine the robustness and ease of use typical of avalanche photodiodes with unprecedented photon number resolution capability, negligible dead time, high detection quantum efficiency and wide sensitive areas. For this reason, they have been successfully exploited in substitution of PMTs in a steadily increasing the number of end-user applications \cite{medical} \cite{cacciaNucl}. Although until now SiPM were sparingly adopted as sensors in biophotonics, the idea that they might eventually emerge as the natural competitors of PMTs also in this field is gaining momentum \cite{CacciaReview} \cite{santangelo}.

SiPM can essentially be operated in two different acquisition modalities: single-photon counting (SPC) and charge integration (CI) \cite{klanner2019}. In SPC, a comparator is used to set a threshold on the detector output. Each time the output current value exceeds the threshold, the event is tagged and the frequency of occurancies is sampled over suitable time bins. In CI, the whole charge from the SiPM signal in a predetermined time window is integrated, measured and eventually digitized to get an arbitrary scale proportional to the number of photons detected in that time interval. These two acquisition modes present complementary advantages and limits. While SPC assures the best performances in terms of ultimate sensitivity, it is affected by saturation at relatively low light intensities due to a phenomenon known as pile-up, consisting in the superposition of single-photon signals. Conversely, baseline instability and non-poissonian noise introduced by the required electronics deteriorate the performances of CI in the low light intensity domain.

In this article, using customized light signals produced by a stochastically triggered, sub-nanosecond pulsed LED source, we characterize the performance of a multi-purpose, portable, user-friendly SiPM based detection system in terms of low limit of detection (LoD) and dynamic range (qualified in terms of response linearity). The system is implemented in order to be capable to perform parallel acquisition in both SPC and CI modes. The benefits introduced in SPC by reshaping of the detector output through a zero-pole cancellation circuit are quantified. Finally, as a proof-of principle demonstration, we use the system to detect the bioluminescence signals emitted by aequorin, a chemiluminescent protein widely exploited as a genetically encoded intracellular calcium sensor, and known for its particularly low photoluminescence efficiency. Namely, we act on gradually diluted samples of lysate of cells expressing aequorin, activating in vitro the luminescent response by means of injection of 2 mM calcium chloride. Exploiting parallel acquisition both in SPC and CI modalities, we obtain linear response to photoluminescent emission over a lysate concentration range spanning three orders of magnitude, with LoD of few kHz. 

\section{Experimental setup}
\label{Sec:Setup}
\begin{figure}[h]
\begin{subfigure}{.5\textwidth}
  \centering
  \includegraphics[width=7cm]{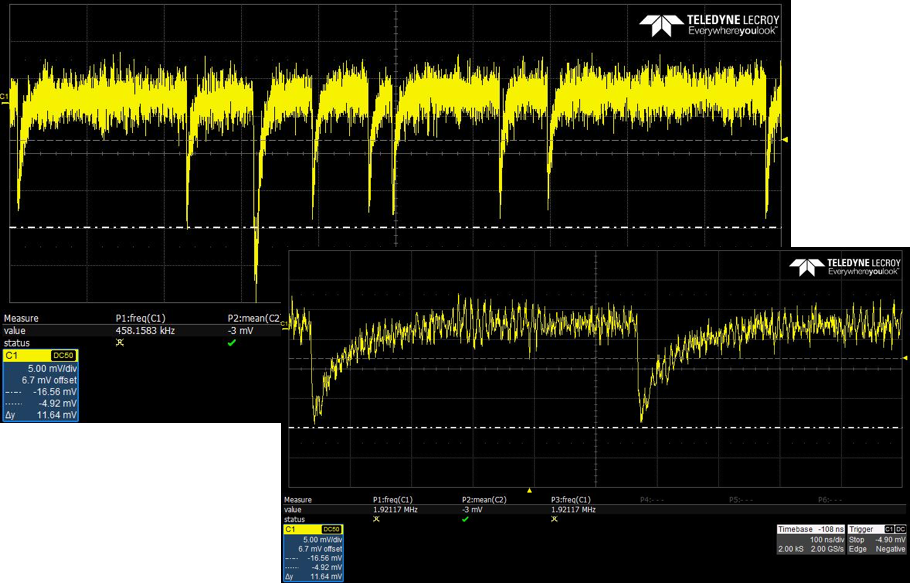}  
  \caption{}
  \label{fig:Nofiltro}
\end{subfigure}
\begin{subfigure}{.5\textwidth}
  \centering
  \includegraphics[width=7cm]{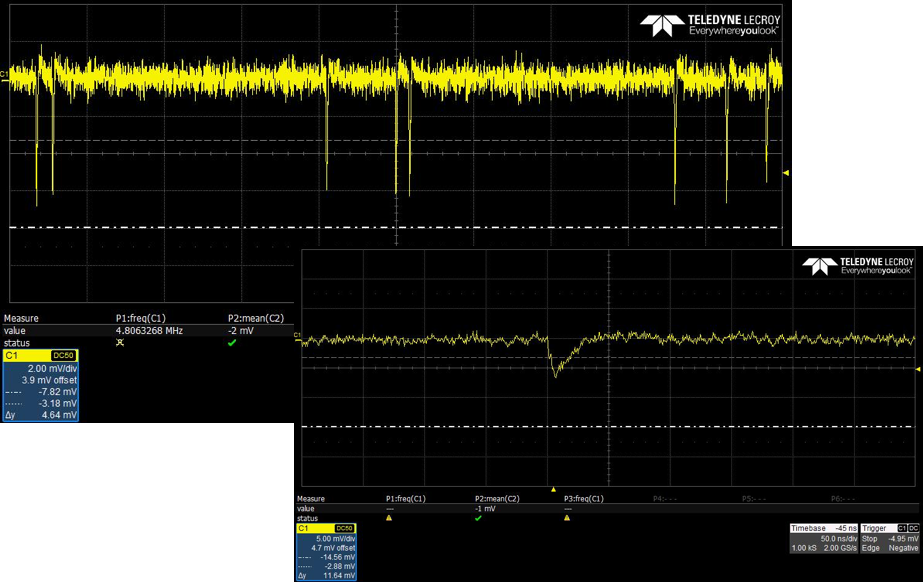}  
  \caption{}
  \label{fig:Filtro}
\end{subfigure}
\caption{(a): the amplified SiPM signal without zero-pole cancellation is shown. (b): the signal after the zero-pole cancellation is presented.}
\end{figure}

The investigation of a system capable of high dynamic range and low sensitivity was performed using a 6$\times$6 mm$^2$ SiPM with 50 $\mu$m pitch (14400 cells in total).
The optimal bias voltage of 55 V ($\sim$+3 overvoltage) is provided by the Power Supply Amplifier Unit (PSAU SP5600 by CAEN S.p.A.). The SiPM was illuminated with a light source produced by a stochastically triggered  LED (PLS-8-2-443 sub-nanosecond-pulsed LED by PicoQuant). The stochastic trigger was produced discriminating the white noise, generated by a Agilent 33250A waveform generator, allowing to explore frequencies ranging from few kHz to some MHz. The SiPM output signal was fed into a custom low noise amplifier. The latter presents two branches with different gain: $\times$50 and $\times$10. The high gain branch is AC coupled (with an output capacitor of 1 $\mu$F), while the low gain one is DC coupled, making the system sensitive to continuous photon fluxes, i.e. to photon frequencies so high that the baseline restoration is not possible.
The full-time development of both output signals is 150 ns (Figure \ref{fig:Nofiltro}). The SPC capability can be enhanced by shortening the signal by means of a zero-pole cancellation circuit. We designed a filter allowing to reshape the output signals to 30 ns full-time development (Figure \ref{fig:Filtro}). 

For the SPC mode tests, the SiPM signal is digitized and discriminated at 0.5-photoelectrons by a Desktop Digitizer DT5730 (14 bit 2V peak to peak), produced by CAEN S.p.A.. The digitizer trigger output (GPIO) is counted over 100 ms by a counter hosted in the PSAU with a sampling rate of 2Hz. 

For the CI mode, the low gain branch amplified signal output was used. The charge integration was performed both by means of an analogue integrator (model V792N, CAEN S.p.A.) and by a Desktop Digitizer DT5720 (CAEN S.p.A.,12 bit, 2V peak to peak, minimum detectable amplitude 0.488mV) through an online proprietary software that digitizes and integrates signals. The sampling rate was set to 35 Hz for both instruments.

\section{Results}
\subsection{Single photon counting modality}
\begin{figure}
\centering
\includegraphics[width=7cm]{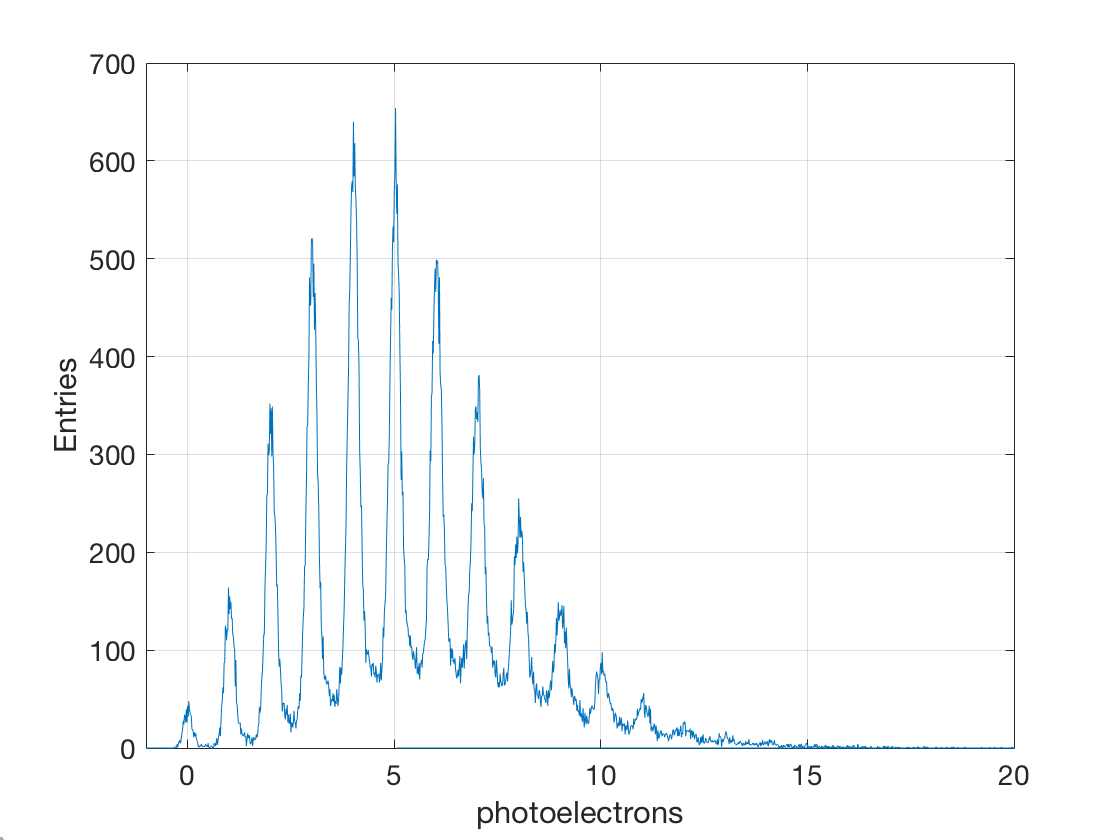}  
\caption{Multiphoton peak spectrum of the typical light used in SPC experiments.}
\label{fig:Multi_high}
\end{figure}

\begin{figure}[h]
\begin{subfigure}{.4\textwidth}
  \centering
  \includegraphics[width=7cm]{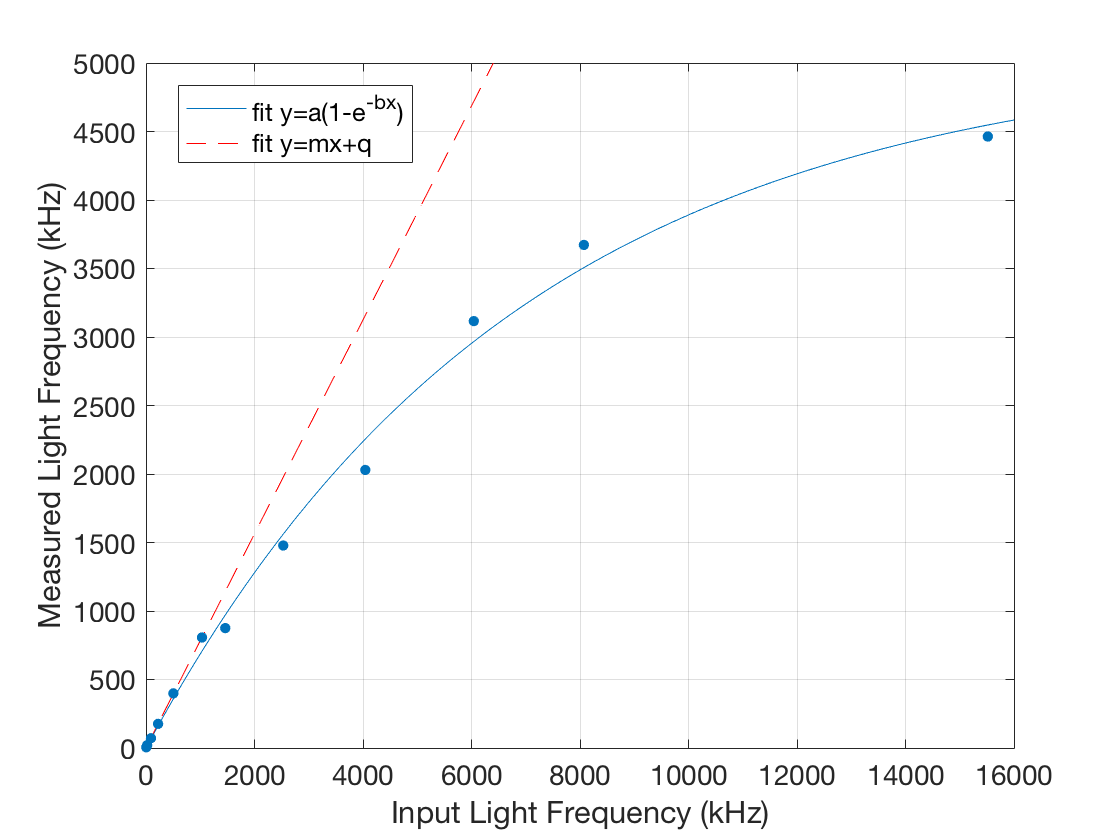}  
  \caption{}
  \label{fig:Counting_noFiltro}
\end{subfigure}
\begin{subfigure}{.4\textwidth}
  \centering
  \includegraphics[width=7cm]{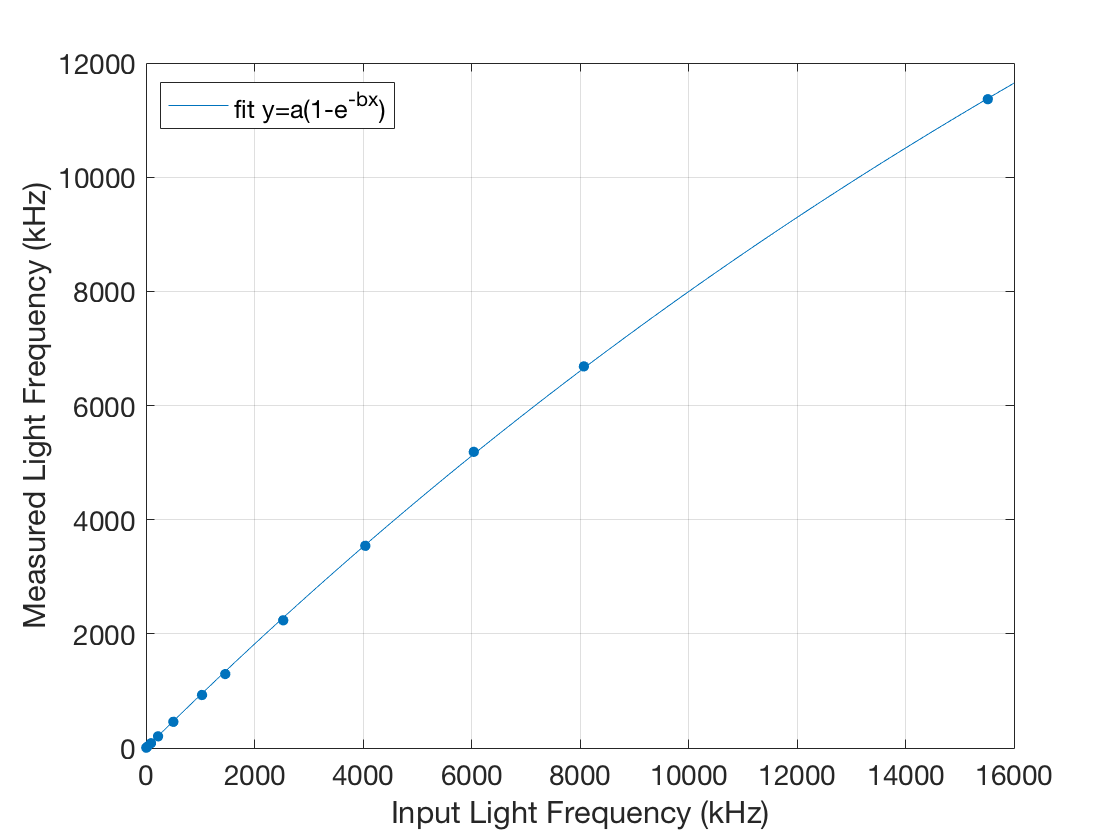}  
  \caption{}
  \label{fig:Counting_Filtro}
\end{subfigure}
\caption{(a): the impinging light frequencies measured as difference of the average count rates with LED-on and LED-off. The blue line represents the fit of the data pointing out a saturation effect. The dashed red line is the linear fit to low frequency points, whose angular coefficient is about 0.78. (b): in this case the SiPM signal output is shortened by means of a pole-zero cancellation circuit. The blue line represents the fit of the data pointing out a saturation effect.}
\label{fig:Multi}
\end{figure}

In order to evaluate the SPC as potential working modality for photoluminescence detection, stochastic LED pulses at average intensity $\sim$5 photoelectrons/pulse (see Figure \ref{fig:Multi_high}  for an exemplary photon number distribution) and increasing mean frequency spanning from few kHz to several MHz were delivered to the SiPM detector through a multimode fibre. This mean value of light assures a null-probability below 1\%. In this way, an improper estimation of the impinging photon frequency is avoided. A dark-count rate (DCR) measurement was repeated for each LED pulse frequency by switching the LED off, and the LED pulse frequency was estimated as the difference between the average count rate detected with LED-on and LED-off. These net frequencies were plotted as a function of the actual LED pulse frequency (see Figure \ref{fig:Counting_noFiltro}). The latter was measured by discriminating the SINC-out NIM signal produced by the LED driver through a scaler (N1145 by CAEN S.p.A). Two deviations from ideal detection were observed:
\begin{itemize}
\item  A sublinear increase above 500 kHz, which indicates a tendency to saturation ascribable to pile-up;
\item A slope of 0.78 $\pm$ 0.01 instead of unit in the linearity range, which shows a substantial underestimation of the LED pulse frequency even when the above phenomena should be negligible.
\end{itemize}

We first investigated the saturation phenomenon. The data in Figure \ref{fig:Counting_noFiltro} were fit to the following equation:
\begin{equation}
y=a\cdot(1-e^{-b\cdot x})
\label{saturazioneFormula}
\end{equation}
In Eq.\ref{saturazioneFormula} the parameter $a$ indicates the saturation frequency, which in the present instance results to be $\nu_{sat}$= (5.0$\pm$0.2) MHz, compliant with a full-time signal development of 150 ns. 
The pile-up probability depends on the total (signal plus DCR) count rate $\nu$ and on the detector output pulse duration $\tau$.
Particularly, the linearity range in SPC can be extended to higher $\nu$ if $\tau$ is reduced. In order to evaluate the benefit deriving from this strategy, we reproduced the measurements of Figure \ref{fig:Counting_noFiltro} reshaping the analogic output signal by means of a custom-made zero-pole cancellation circuit to a full-development duration of the single-photon pulse of $\sim$30 ns. The results are plotted in Figure \ref{fig:Counting_Filtro}, and assert that the linearity of the system is significantly improved by this implementation. Indeed, fitting this dataset to Eq. \ref{saturazioneFormula} yields a saturation frequency value as high as $\nu_{sat}$= (27.9$\pm$0.6) MHz. It is worth mentioning that zero-pole cancellation circuits allowing to shorten an income signal similar to our rough analogic output pulses to as few as 5 ns are currently available \cite{polacchi}.  

\begin{figure}
\centering
\includegraphics[width=7cm]{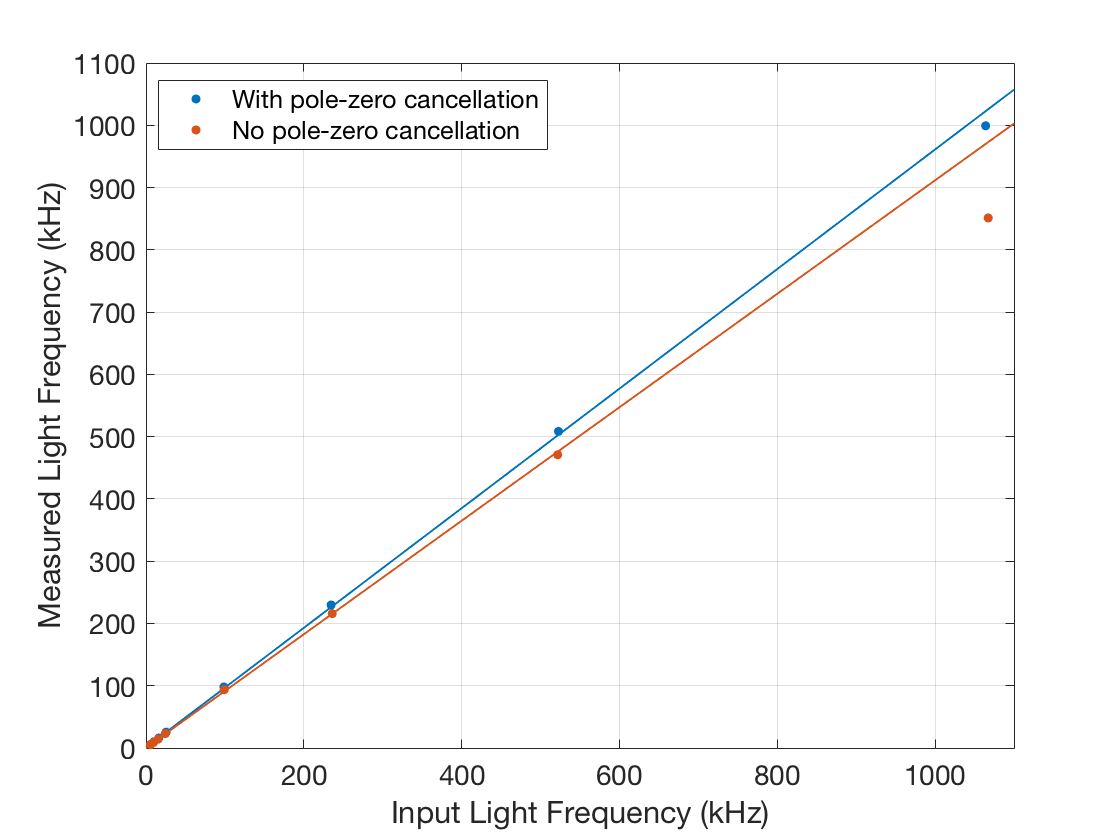}  
\caption{The coincidence count rates measured with (blue dots) and without (red dots) the zero-pole cancellation circuit at frequency below 1MHz. The two lines represent the linear fits.}
\label{fig:Slope}
\end{figure}

As the detector has non-negligible DCR (800 $\pm$ 3 kHz), we hypothesized that the underestimation of LED pulses at low frequencies might be due to simultaneous arrival of DCR with light pulses, which corresponds to a net reduction in digitizer output pulses generated by spurious DCR pulses with respect to the LED-off situation, and results in a systematic error in baseline subtraction. This hypothesis was tested by recurring to coincidence measurements in which a NIM signal of pulse duration 100 ns, synchronous with the LED pulses, was correlated with the digitizer output exploiting a N405 logic unit (CAEN S.p.A.). The above pulse duration was chosen as an estimate of the time required for the analogic SiPM output signal to return below threshold after a detection event. It is worth to recall that this time depends on the pulse height, thus ultimately on the intensity of the detected laser pulses. We decided to fix the pulse duration value to that required for a 5-photoelectrons pulse to return below threshold, as this was the average intensity of our light source. Because in coincidence modality a detector output pulse is counted as a light detection event only if it is synchronous with a light pulse, DCR subtraction is unnecessary, and the rate of coincidences should approach the input light pulse frequency when pile-up and baseline shift effects are negligible. Indeed, as shown in Figure \ref{fig:Slope} the slope of the linear range of the coincidence rate versus LED pulse frequency plot is 0.91 $\pm$ 0.01, which indicates an almost ideal detection efficiency below $\sim$1 MHz, taking into account that our discriminator is able to disentangle two events only if they are temporally separated by more than $\sim$125 ns. The probability of losing events within 125 ns is $1-e^{-\mu}$, where $\mu$ represents the expectation value of counts within this window. For a LED pulse frequency of 100 kHz, in the middle of the linearity range, this corresponds to $\sim$10\%. This interpretation is supported by the observation that, performing similar measurements with the analogic detector output reshaped by the zero-pole cancellation circuit the slope in the linearity range increases to 0.96 $\pm$ 0.01, due to the fact that in this instance subsequent pulses, being shorter, are on the average more distant in time.

\subsection{Charge integration modality}
\begin{figure}
\centering
\includegraphics[width=6.5cm]{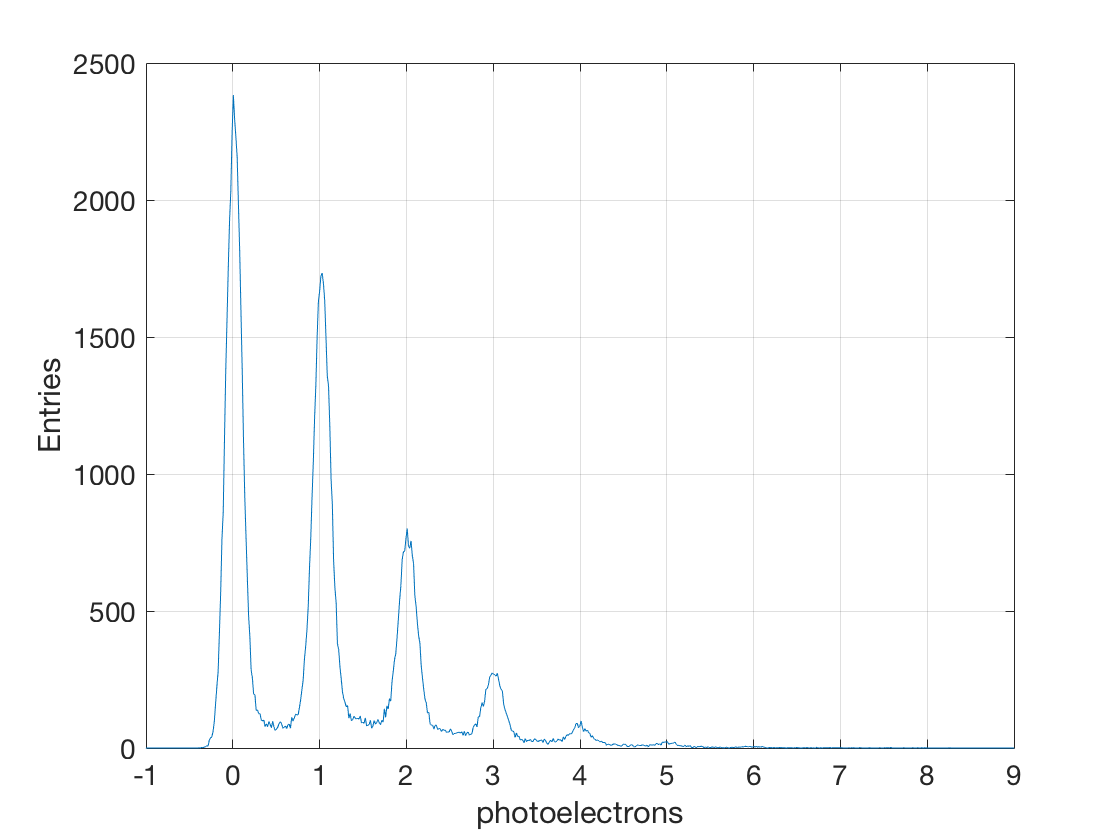}  
\caption{Multiphoton peak spectrum of the typical light used in CI experiments.}
\label{fig:Multi_low}
\end{figure}

\begin{figure}[h]
\begin{subfigure}{.4\textwidth}
  \centering
  \includegraphics[width=7cm]{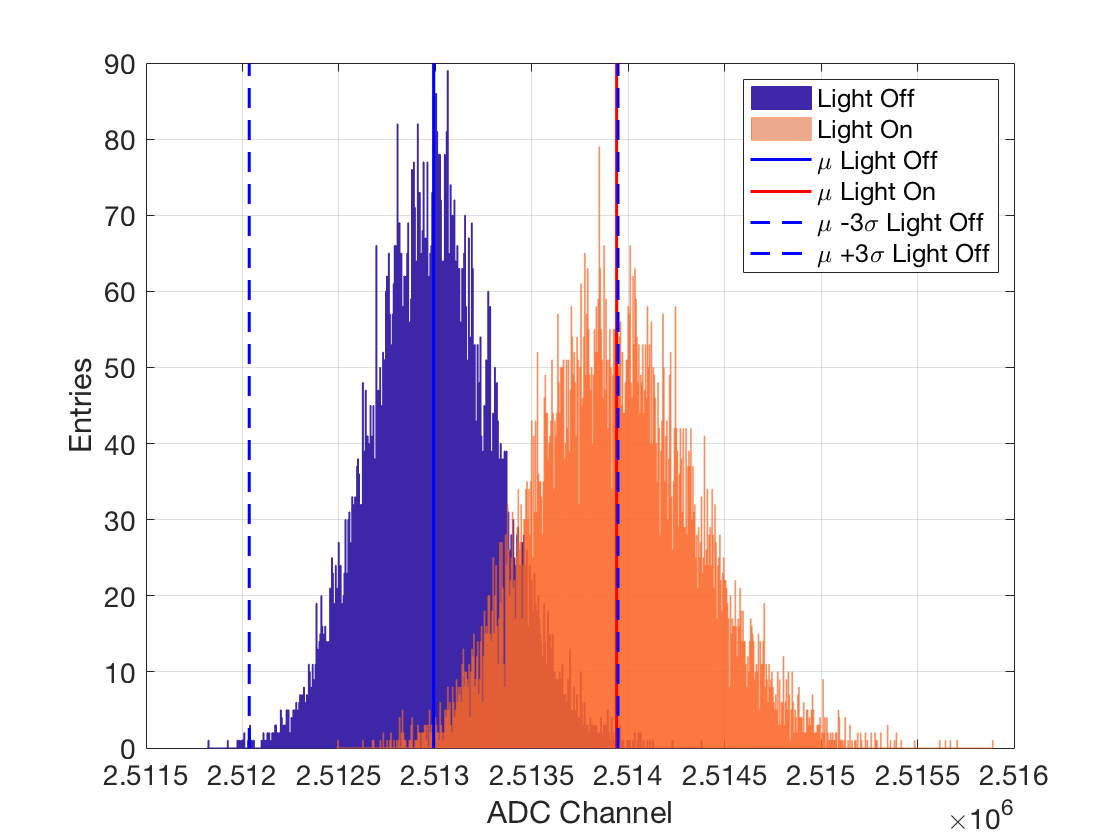}  
\end{subfigure}
\begin{subfigure}{.4\textwidth}
  \centering
  \includegraphics[width=7cm]{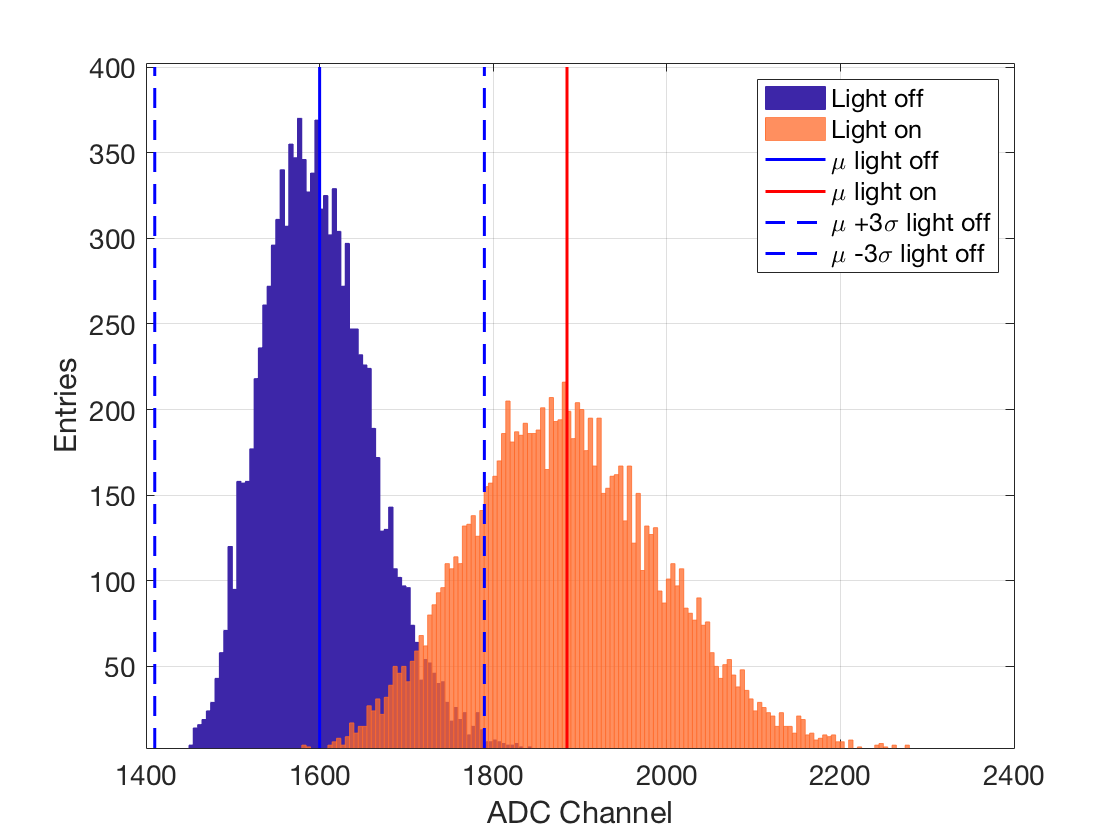}  
\end{subfigure}
\caption{(a): the charge distribution in CI with the digitizer: in blue the dark current of the biased SiPM (amplifier noise + SiPM DCR), the blue line is the mean value ($\mu$), while the two dashed lines are the limit at 3 $\sigma$ in an integration gate of 5 $\mu$s. In orange the charge distribution of the light generated at frequency of 2500 kHz. The red line corresponds to its mean value. (b): the same distributions recorded by the QDC system.}
\label{fig:distri}
\end{figure}

In order to compare the CI performances with the SPC ones, the dark current ($\mu_{dark}$) of the biased SiPM and then the current induced by the illumination ($\mu_{light}$) were measured within an integration gate of 5 $\mu$s, as described in section \ref{Sec:Setup}. In this case the mean value of the light was kept at $\sim$1 photoelectron (see Figure \ref{fig:Multi_low}  for an exemplary photon number distribution) in order to mimic the typical experimental conditions of a bioluminescence measurement, in which trains of single photons are emitted. 
The values of $\mu_{dark}$, $\mu_{light}$ and their difference $\Delta=\mu_{light}-\mu_{dark}$ were evaluated with both systems (QDC and Digitizer 5720).The standard deviations of the mean value of each distribution are added in quadrature to compute the error on the difference. Examples of the obtained distributions are reported in Figure \ref{fig:distri}.

\begin{figure}[h]
\begin{subfigure}{.4\textwidth}
  \centering
  \includegraphics[width=7cm]{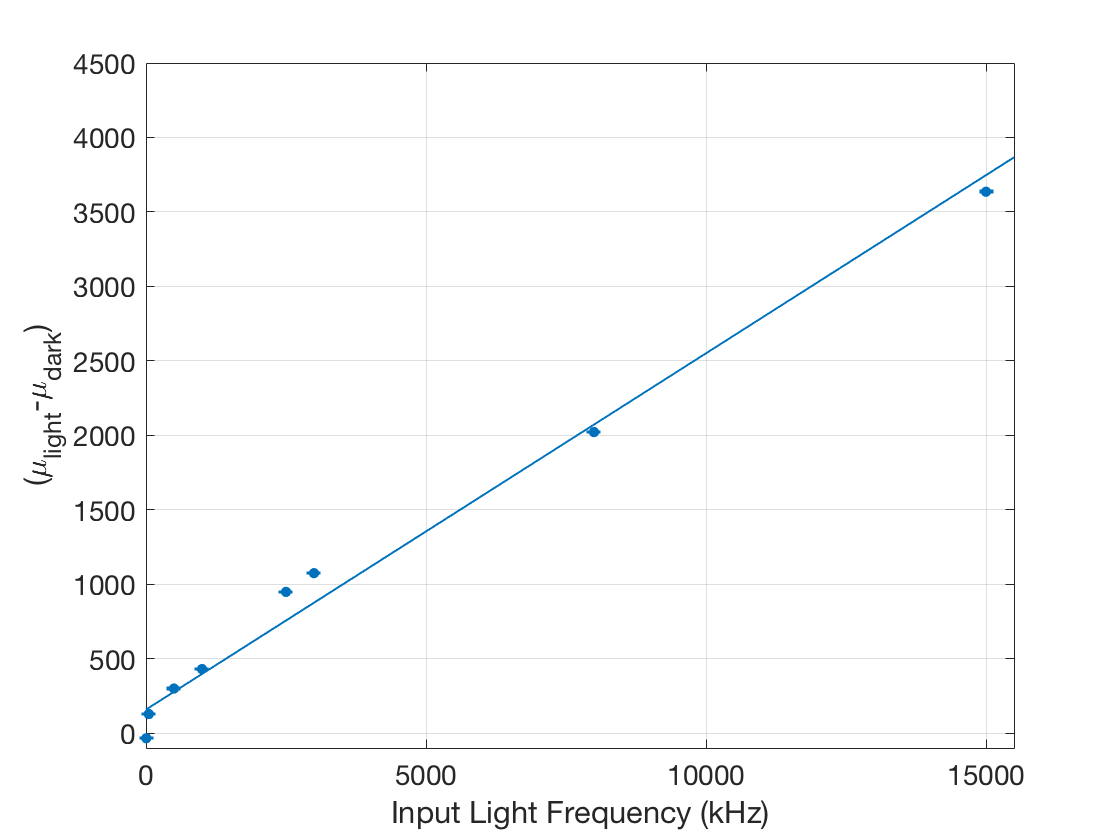}  
\end{subfigure}
\begin{subfigure}{.4\textwidth}
  \centering
  \includegraphics[width=7cm]{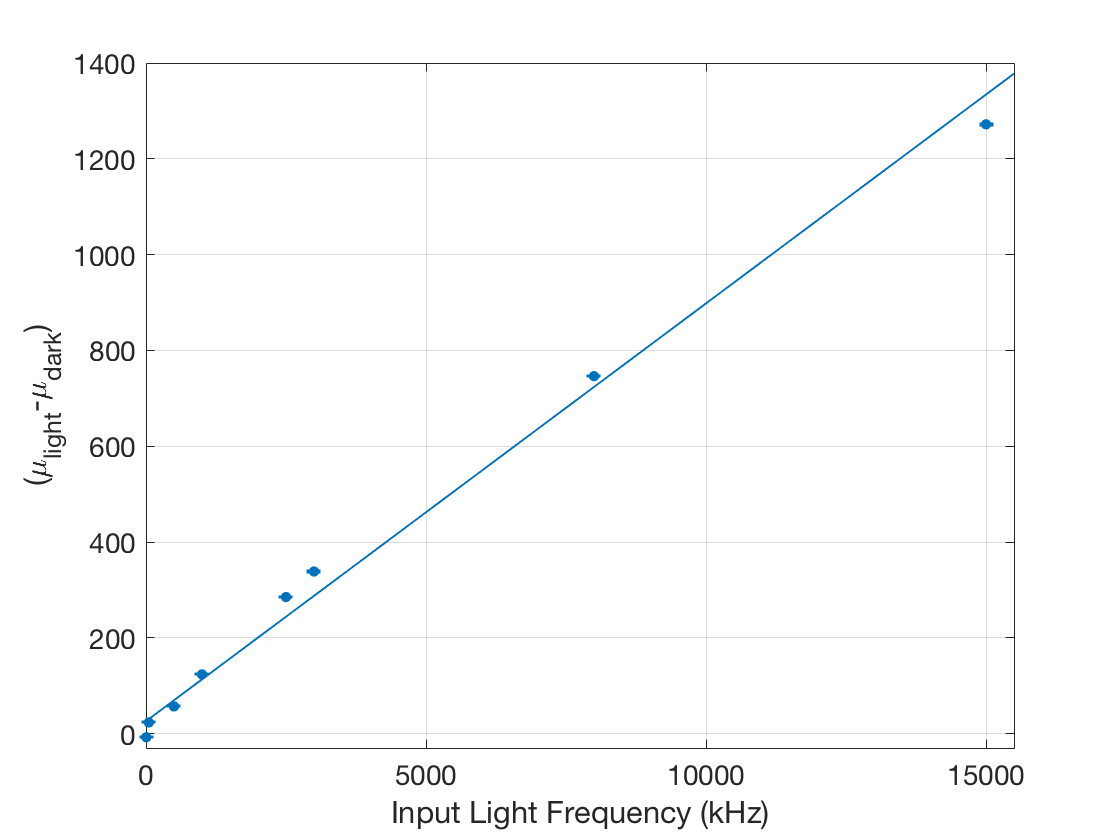}  
\end{subfigure}
\caption{(a): Variations of the integrated charge with and without the impingin light on the sensor as a function of the impinging photons frequency for the Digitizer DT5720. The line represents the fit interpolating all the points. (b): the same for the analog integrator QDC}
\label{fig:Lin_mean1}
\end{figure}

As shown in Figure \ref{fig:Lin_mean1}, with both devices the plot of $\Delta$ values as a function of LED pulse frequency is roughly linear over more than 4 orders of magnitude. As expected, the intercepts are compatible with zero within a 2$\sigma$ confidence interval. 
Linear coefficients of m=(0.24 $\pm$ 0.02) ADC/kHz and m=(0.087$\pm$ 0.005) ADC/kHz where determined for the Digitizer and QDC, respectively. This results configure the Digitizer as more sensitive than the QDC by a factor 3. An important figure of merit is the limit of detection (LoD), expressed in terms of the minimal frequency which can be discriminated from noise. We define the LoD as $LoD=\Delta/\sigma_{dark}\geq3$, where $\sigma_{dark}$ is the standard deviation of the noise distribution and were measured to be 321 ADC for the Digitizer and 67 ADC for the QDC. Accordingly, the LoD corresponds to $\sim$4MHz for the Digitizer and to $\sim$2.3 MHz for the QDC. The higher value with the digitizer can be explained considering that on the average the digitized signals correspond to single photons with a peak amplitude of $\sim$3 mV and that the minimum amplitude that can be digitized is 0.488 mV. 
Consequently, only few points are above the noise and higher fluctuations in the integrated charge are expected. Moreover, it should be noticed that this intrinsic limitation is higher than the expected value if a poissonian noise is considered. Indeed, in case of a poissonian noise the LoD$_{poiss}$ can be described as 
\begin{equation}
LoD_{poiss}\geq\frac{3\sqrt{DCR\cdot \Delta t}}{\Delta t}
\end{equation}
that would lead to a LoD of $\sim$ 380 kHz considering a DCR of 800 kHz and a $\Delta t$ of 5 $\mu$s. Since the LoD is higher for both our system, a motivation could be that a non-poissonian noise is induced by the amplification system. The LoD can be reduced exploiting the higher sampling rate of CI systems (35 Hz) with respect of the SPC one (2Hz) and averaging over N consecutive events. Taking N$=$17 in our case the LoD decreases to 1.2 MHz for digitizer and to 500 kHz for QDC. These LoD values should be compared with the one expected (and actually measured) for the SPC system assuming a poissonian distribution of noise and a DCR of 800 kHz, i.e. LoD=3$\sigma_{DCR}$=9 kHz. Thus, SPC is has superior ultimate sensitivity with respect to CI, although it is plagued by a smaller linearity range due to pile-up.

\subsection{Proof of principle application to detection of a real bioluminescence signal}
In the sought of providing a preliminary assessment of the performances of our SiPM-based luminometry system in the measurement of real luminescence signals, a lysate of HeLa cells transfected to express cytosolic aequorin was used. The transfection protocol is fully described elsewhere \cite{Lim2016}. Cell lysis was pursued by disrupting the cytoplasmic membranes by repeated freeze-thawing (3$\times$) in the following reaction medium: 150 mM Tris buffer, added with 0.8 mM phenylmethylsulfonyl fluorid and 0.1 ethylenediaminetetraacetic acid. The soluble cell constituents were separated from organelles and membrane phospholipids by centrifugation at 12000 rpm for 5 min at 4$^\circ$C. The supernatant, containing aequorin, was reconstituted overnight on ice by addition of 5 $\mu$M coelenterazine, the biological substrate of this chemiluminescent protein, and 140 mM $\beta$-mercaptoethanol. 

\begin{figure}[h]
\begin{subfigure}{.4\textwidth}
  \centering
  \includegraphics[width=7cm]{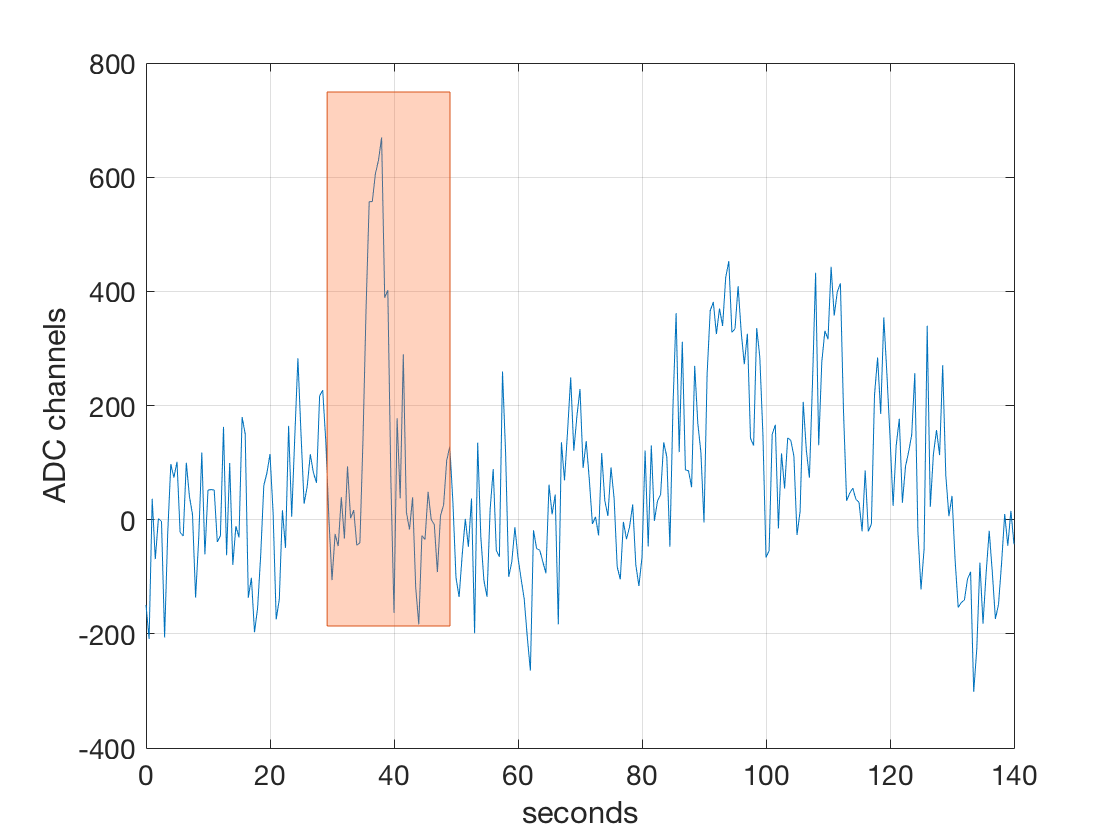}  
\end{subfigure}
\begin{subfigure}{.4\textwidth}
  \centering
  \includegraphics[width=7cm]{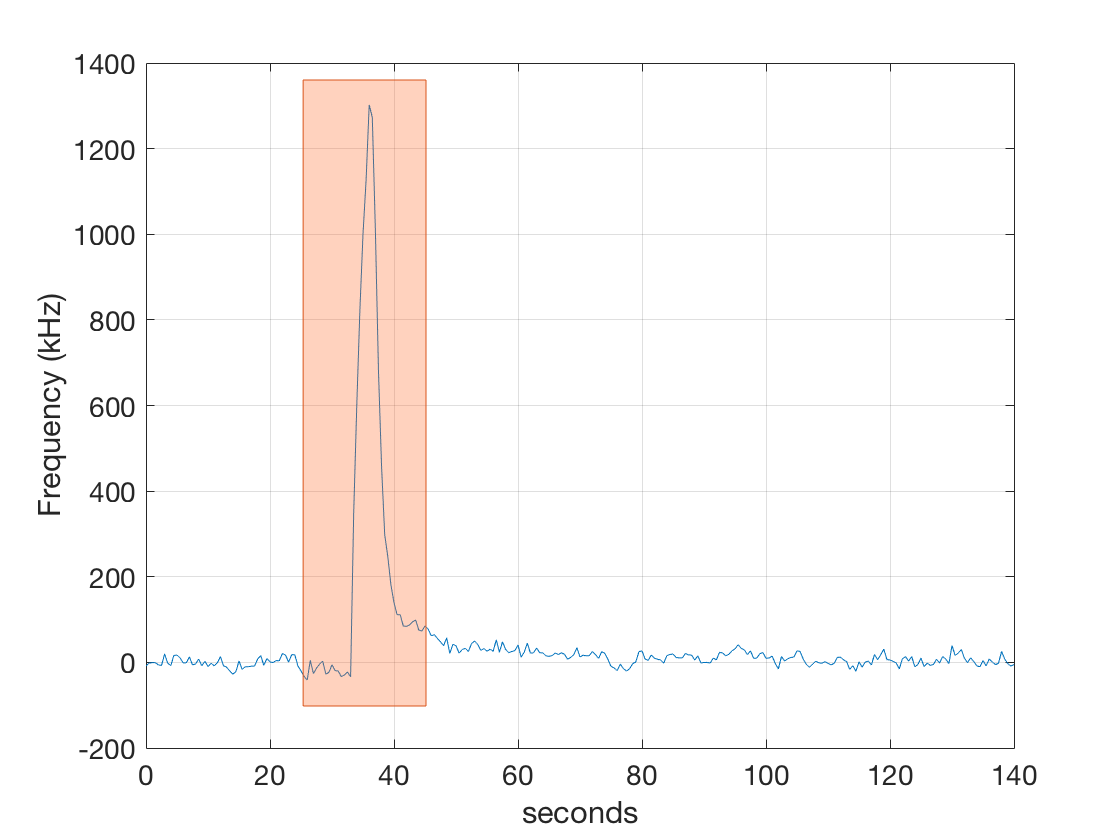}  
\end{subfigure}
\caption{(a): recorded trace of the bioluminescent signal using the CI acquisition mode with the digitizer in correspondence of a dilution factor of 7 times (the lowest detectable for this system). The orange box highlights the integral intensity of the bioluminescent signal which convey the information of interest in this experiments. (b): the same trace recorded simultaneously with the SPC modality.}
\label{fig:calciumSpike}
\end{figure}

For the bioluminescence detection experiment, the reconstituted cell lysate was progressively diluted in 150 mM Tris-buffer (pH 7.2). For each concentration, 0.5 ml of lysate were put in a perfusion chamber situated directly on the detector sensitive area. Optical grease was used in order to optimize the refraction index matching between chamber and detector. The detector and chamber were kept in the dark. The detector output signal was split and sent in parallel to the PSAU counting unit, through the high-gain channel of the amplifier, and to the digitizer DT5720, through the low-gain channel of the amplifier, in order to process the data in SPC and CI modes simultaneously. After detection of the basal light signal for 1 min, 0.5 ml of CaCl$_2$ 2 mM were injected into the chamber to elicit the chemiluminescent response. Exemplary traces for both acquisition system are reported in Figure \ref{fig:calciumSpike} for a Aequorin concentration of 32 A.U, i.e the nominal highest concentration equal to 4096 A.U. was diluted 7 times.

\begin{figure}[h]
\begin{subfigure}{.4\textwidth}
  \centering
  \includegraphics[width=7cm]{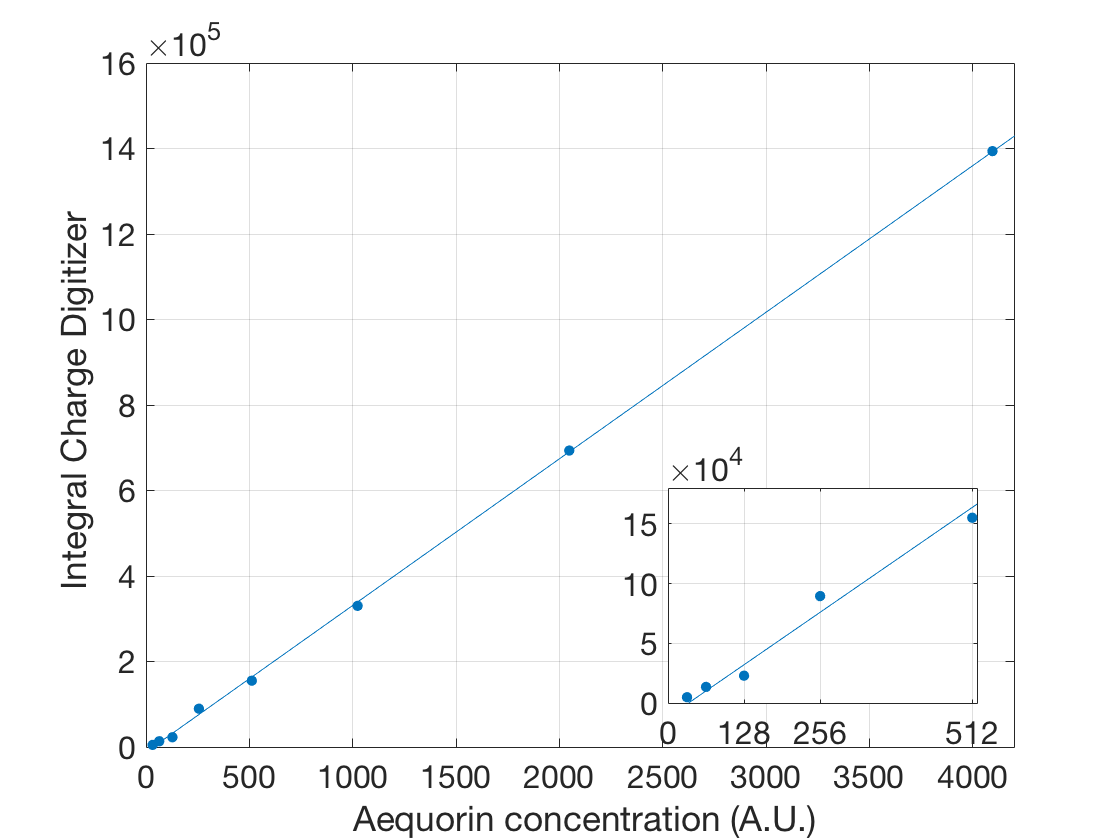}  
\end{subfigure}
\begin{subfigure}{.4\textwidth}
  \centering
  \includegraphics[width=7cm]{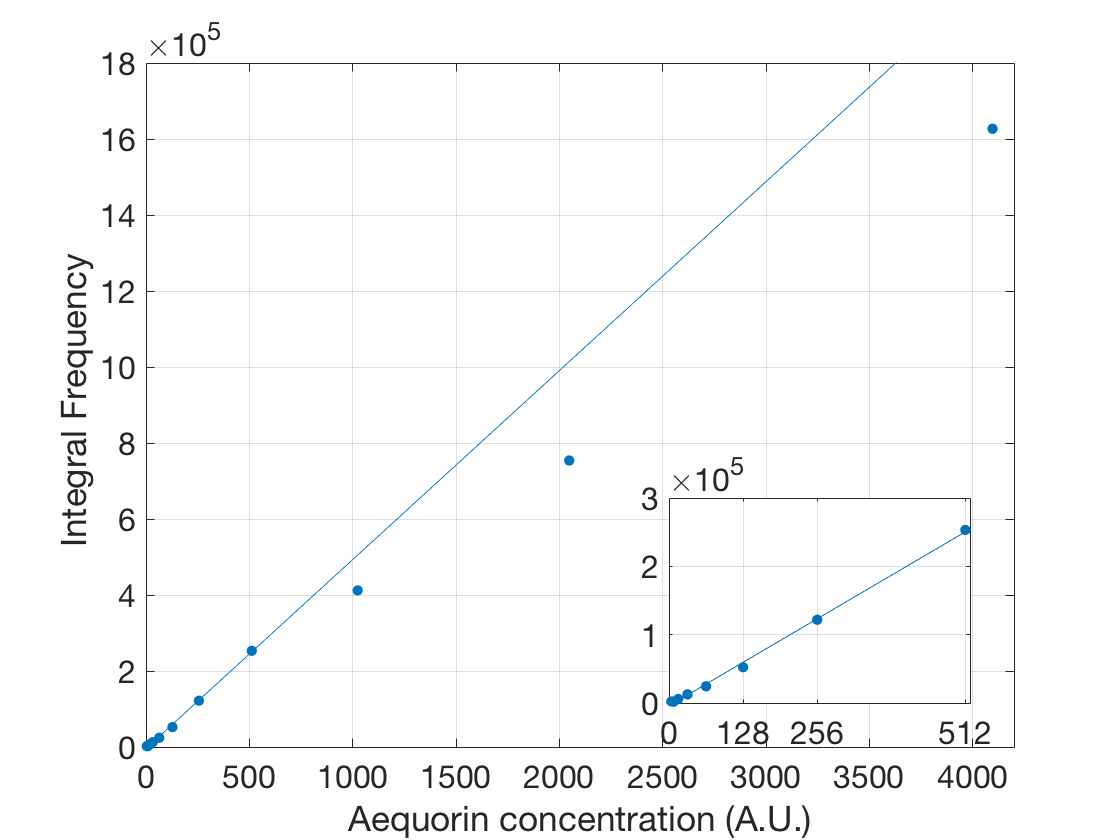}  
\end{subfigure}
\caption{(a): the integrated intensity of the bioluminescent signal is represented as a function of Aequorin concentration in A.U, obtained in CI mode with digitizer. The line represents the fit over all the data points. (b): the same plot obtained in SPC modality. In this case a deviation from linearity is shown at high concentration where probably the pile-up phenomenon affects so many points of the trace that also the integral of the Ca$^{2+}$ pulse is plagued by this phenomenon.}
\label{fig:calciumLin}
\end{figure}

First of all, thanks to the parallel acquisition the two bioluminescence responses are directly comparable, once the CI recorded data are averaged over 17 events in order to keep the same sampling frequency of 2 Hz used in SPC. As shown in Figure \ref{fig:calciumSpike}, when the frequency of the photons emitted by the biological sample is below 1.2 MHz the charge integrated results in a very feeble signal above the noise, confirming the LoD obtained in the lab characterization. 
Conversely, the SPC response is clearly distinguishable from noise until the nominal Aequorin concentration of 2 A.U., that is over more than a further order of magnitude in dilution. 
It is worth mentioning that in this experiment the biological information of the process lies in the integral of the induced signal  (orange box in Figure \ref{fig:calciumSpike}) and is expected to increase with the concentration of aequorin in lysate. Exploiting the parallel acquisition in SPC and CI, it is possible to detect bioluminescent signals that range over 3 order of magnitude in terms of lysate concentration, as shown in Figure \ref{fig:calciumLin}.

\section{Conclusions}
In this Article the performances of a portable bioluminescence detection system based on a SiPM detector and a multi-purpose system for the signal readout were assessed in terms of response linearity range and ultimate sensitivity (lower limit of detection) using customized light signals generated by a randomly-triggered pulsed LED. A proof of principle demonstration of the system ability to detect real bioluminescence signals was pursued on a simplified in vitro model system consisting in a lysate of cells expressing the widespread intracellular calcium sensor aequorin. The biosensor response was elicited by injection of calcium chloride at high concentration. Parallel acquisition in single-photon counting and charge integration modes allowed to detect the aequorin chemiluminescence signal over a biosensor concentration range spanning more than three orders of magnitude, with ultimate sensitivity down to few kHz. The systematized characterization allowed to identify the figures of merit of the detector and readout circuitry on which engineering activities should concentrate in order to optimize the performance of the system as a luminometer. Namely, reshaping of the analogic output signal to shorten the full-development pulse duration and reducing the dark-count rate promise to result particularly beneficial implementations.
\section*{References}

\end{document}